\pdfoutput=1
\newcommand*{\ATLASLATEXPATH}{latex/}
\documentclass[cernpreprint, texlive=2016, UKenglish]{\ATLASLATEXPATH atlasdoc}
\usepackage{\ATLASLATEXPATH atlaspackage}
\usepackage{\ATLASLATEXPATH atlasbiblatex}
\usepackage{\ATLASLATEXPATH atlasphysics}

\usepackage{HIGG-2017-09-PAPER-defs}

\AtlasTitle{Search for Higgs boson decays into pairs of light (pseudo)scalar particles in the $\gamma\gamma jj$ final state in $pp$ collisions at $\sqrt{s}=13$ \TeV{} with the ATLAS detector}

\AtlasAbstract{%
This Letter presents a search for exotic decays of the Higgs boson to a pair of new (pseudo)scalar
particles, $H\to aa$, where the $a$ particle has a mass in the range 20--60 \GeV{}, and where one of the $a$ bosons decays into a pair of photons and the other to a pair of gluons.
The search is performed in event samples enhanced in vector-boson fusion Higgs boson production
by requiring two jets with large invariant mass in addition to the Higgs boson candidate decay products.
The analysis is based on the full dataset of $pp$ collisions at $\sqrt{s}= 13$ \TeV{} recorded in 2015 and 2016 with the ATLAS detector at the CERN Large Hadron Collider, 
corresponding to an integrated luminosity of 36.7 \ifb. 
The data are in agreement with the Standard Model predictions and an upper limit at the 95\% confidence level is placed on the production cross section times the 
branching ratio for the decay $H\to aa\to \gamma\gamma gg$. 
This limit ranges from 3.1 pb to 9.0 pb depending on the mass of the $a$ boson.
}

 \author{The ATLAS Collaboration}

\AtlasRefCode{HIGG-2017-09}

\PreprintIdNumber{CERN-EP-2017-295}

 \AtlasJournal{Phys.\ Lett.\ B.}
 \AtlasJournalRef{\PLB 782 (2018) 750}
 \AtlasDOI{DOI:10.1016/j.physletb.2018.06.011}

 \AtlasCoverSupportingNote{Supporting note}{https://cds.cern.ch/record/2259309}

\AtlasCoverCommentsDeadline{18 March 2018}

\AtlasCoverEgroupEditors{atlas-higg-2017-09-editors@cern.ch}

\AtlasCoverEgroupEdBoard{atlas-higg-2017-09-editorial-board@cern.ch}

\hypersetup{pdftitle={ATLAS document},pdfauthor={The ATLAS Collaboration}}

\begin{document}

\maketitle

\section{Introduction}

The discovery or exclusion of the Standard Model (SM) Higgs 
boson was one of the main goals of the Large Hadron Collider (LHC) physics programme.  
A Higgs boson with mass  of 125~\gev, and with  properties compatible with those expected for the SM Higgs boson ($H$), 
was discovered by the ATLAS~\cite{HIGG-2012-27} and CMS~\cite{CMS-HIG-12-028} collaborations.
Since its discovery, a comprehensive programme of measurements of the properties of this particle has been underway.
These measurements could uncover deviations from branching ratios predicted by the SM or 
set a limit on the possible branching ratio for decays into new particles beyond the SM (BSM).
Existing measurements constrain the branching ratio for such decays ($B_\text{BSM}$) 
to less than 34\% at 95\% confidence level (CL)~\cite{HIGG-2015-07}, 
assuming that the absolute couplings to vector bosons are smaller than or equal to the SM ones.

Many BSM models predict exotic decays of the Higgs boson~\cite{Curtin:2013fra}.
One possibility is that the Higgs boson decays into a pair of 
new (pseudo)scalar particles, $a$, which in turn decay to a pair of SM particles.
Several searches have been performed for $H\to a a$ in various final 
states~\cite{Abazov:2009yi,CMS-HIG-13-010,CMS-HIG-16-015,EXOT-2013-15,HIGG-2014-02}.

The results presented in this Letter cover the unexplored $\gamma\gamma{jj}$ final state in searches for
$H\to a a$, where one of the $a$ bosons decays into a pair of photons and the other decays into a pair of gluons.
This final state becomes relevant in models where the fermionic decays are suppressed
and the $a$ boson decays only into photons or gluons~\cite{Curtin:2013fra,hep-ph/0703247}.
The ATLAS Run 1 search for $H\to aa \to \gamma\gamma\gamma\gamma$~\cite{EXOT-2013-24} set
a 95\% CL limit $\sigma_H\times B(H\to aa\to \gamma\gamma\gamma\gamma)<10^{-3}~\sigma_\text{SM}$
for $10<m_a<62~\GeV{}$, where $\sigma_\text{SM}$ is the production cross-section for the SM Higgs boson.
There is currently no direct limit set on $B(H\to aa\to \gamma\gamma gg)$;
however, in combination with $B_\text{BSM}<34\%$,
the $H\to aa \to \gamma\gamma\gamma\gamma$ result sets an indirect limit on $B(H\to aa\to \gamma\gamma gg)$ to less than $\sim4\%$.
Assuming the same ratio of photon and gluon couplings to the $a$ boson as to the SM Higgs boson,
the $H\to aa \to \gamma\gamma\gamma\gamma$ decay occurs very rarely relative to the $H\to aa \to \gamma\gamma gg$ decay 
(a typical value for the ratio $B(H\to aa\to \gamma\gamma\gamma\gamma)/B(H\to aa \to \gamma\gamma gg)$ is $3.8\times10^{-3}$~\cite{hep-ph/0703247})
making $H\to aa \to \gamma\gamma jj$ an excellent unexplored final state for probing these fermion-suppressed coupling models.
The branching ratio for $a\to \gamma\gamma$ can be enhanced in some scenarios.
The two searches are therefore complementary,
where the $H\to aa\to \gamma\gamma jj$ final state is more sensitive to photon couplings
with the new physics sector similar to the photon coupling to the SM Higgs boson,
while the $H\to aa\to \gamma\gamma\gamma\gamma$ final state is more sensitive to scenarios with enhanced photon couplings.
In addition, the $H\to aa\to \gamma\gamma jj$ final state can probe models inaccessible by the $H\to aa\to \gamma\gamma\gamma\gamma$  final state, for example $H\to aa^\prime\to\gamma\gamma jj$ where the $a$ and $a^\prime$ are both (pseudo)scalar particles with similar masses with primary decays to photons and gluons, respectively.

Reference~\cite{hep-ph/0703247} shows that the search for $H \to \gamma\gamma gg$, 
where the Higgs boson is produced in association with a vector boson which decays leptonically, 
would require approximately $300~\ifb$ of LHC data in order to be sensitive to branching ratios less than 4\%. 
The gluon--gluon fusion (ggF) production mode has a larger cross-section, 
but is overwhelmed by the $\gamma\gamma$+multi-jet background.
The strategy described in this Letter consists in selecting events where vector-boson fusion (VBF) is the dominant Higgs boson production mode.
Even though the production rate is lower than that for the ggF mode, 
the characteristic topology of the jets produced in association with the Higgs boson enables 
more effective suppression of the background.

\section{Data and simulation}

The search presented in this Letter is based on the 36.7~\ifb{} dataset of proton--proton collisions
recorded by the ATLAS experiment at the LHC at $\sqrt{s}=13$ \TeV{} during 2015 and 2016.
The ATLAS detector~\cite{PERF-2007-01} comprises an inner detector in a 2 T axial magnetic field, 
for tracking charged particles and a precise localisation of the interaction vertex, 
a finely segmented calorimeter, a muon spectrometer and a two-level trigger~\cite{TRIG-2016-01} that
accepts events at a rate of about 1 kHz for data storage.

Monte Carlo (MC) event generators were used to simulate the $H\to aa \to \gamma\gamma gg$ signal.
Signal samples for the ggF and VBF processes were generated at next-to-leading order using 
\POWHEGBOX{}~\cite{Nason:2004rx,Frixione:2007vw,Alioli:2010xd} interfaced with \PYTHIA{}~\cite{Sjostrand:2014zea} for parton showering and hadronisation using the AZNLO set of tuned parameters set~\cite{STDM-2012-23} and the CT10 parton distribution function (PDF) set~\cite{Lai:2010vv}.
Samples were generated in the $m_a$ range\footnote{The diphoton triggers considered for this
search do not have acceptance for the lower mass range ($m_a<20~\GeV{}$), where the two photons are collimated.} 
$20~\GeV{} < m_a < 60~\GeV{}$, assuming the $a$ boson to be a (pseudo)scalar.
All MC event samples were processed through a detailed simulation~\cite{SOFT-2010-01} of the ATLAS detector
based on \geantFour{}~\cite{Agostinelli:2002hh}, and contributions from additional $pp$ interactions (pile-up), simulated using
\PYTHIA{} and the MSTW2008LO PDF set~\cite{Martin:2009iq}, were overlaid onto the hard-scatter events.

\section{Selection criteria}

Events are selected by two diphoton triggers.
One trigger path requires the presence in the electromagnetic (EM) calorimeter of two clusters of energy
deposits with transverse energy\footnote{ATLAS uses a right-handed coordinate system with its origin 
  at the nominal interaction point (IP) in the centre of the detector and the $z$-axis along the beam pipe. The $x$-axis points from the IP to the centre of the LHC ring, 
  and the $y$-axis points upward. Cylindrical coordinates $(r,\phi)$ are used in the transverse plane, $\phi$ being the azimuthal angle around the $z$-axis. 
  The pseudorapidity is defined in terms of the polar angle $\theta$ as $\eta=-\ln\tan(\theta/2)$.}
above 35 \GeV{} and 25 \GeV{} for the leading (highest transverse energy) and sub-leading
(second-highest transverse energy) clusters, respectively. In the high-level trigger the shape of the energy
deposit in both clusters is required to be loosely consistent with that expected from an EM shower initiated by a photon.
The other trigger path requires the presence of two clusters with transverse energy above 22 \GeV{}.
In order to suppress the additional rate due to the lower transverse energy threshold, the shape requirements for the energy deposits are
more stringent.

The photon candidates are reconstructed from the clusters of energy deposits in the EM calorimeter within the range $|\eta|<2.37$.
The energies of the clusters are calibrated to account for energy losses upstream of the calorimeter 
and for energy leakage outside the cluster, as well as other effects due to the detector geometry and response.
The calibration is refined by applying $\eta$-dependent correction factors of the order of $\pm1$\%, derived from $Z\to ee$ events~\cite{PERF-2013-05}.
As in the trigger selection, photon candidates are required to satisfy a set of identification criteria based on the shape of the EM cluster~\cite{PERF-2013-04}.
Two working points are defined: a \textit{Loose} working point, used for the preselection and the data-driven background estimation, and a 
\textit{Tight} working point, with requirements that further reduce the misidentification of neutral hadrons decaying to two photons.
In order to reject the hadronic jet background, photon candidates are required to be isolated from any other activity in the calorimeter.
The calorimeter isolation is defined as the sum of the transverse energy in the
calorimeter within a cone of \mbox{$\Delta R = \sqrt{(\Delta\eta)^2+(\Delta\phi)^2}=0.4$} centred around the photon candidate,
The transverse energy of the photon candidate is subtracted from the calorimeter isolation.
Contributions to the calorimeter isolation from the underlying event and pile-up are subtracted using the method proposed in Ref.~\cite{jetareas}.
Candidates with a calorimeter isolation larger than 2.2\% of the photon's transverse energy are rejected.

Jets are reconstructed from topological clusters~\cite{PERF-2014-07} using the anti-$k_t$
algorithm~\cite{Cacciari:2008gp} implemented in the FastJet package~\cite{Cacciari:2011ma} with a radius parameter of $R=0.4$. 
Jets are calibrated using an energy- and $\eta$-dependent
calibration scheme, and are required to have a transverse momentum (\pt{}) greater than $20~\GeV{}$ and $|\eta|<2.5$ or $\pt>30~\GeV{}$ and $|\eta|<4.4$.
A track- and topology-based veto~\cite{PERF-2014-03,PERF-2016-06} is used to suppress jets originating from pile-up interactions.
Jets must have an angular separation of $\Delta R>0.4$ from any \textit{Loose} photon candidate in the event.

\begin{figure}[t]
  \centering 
  \subfloat[]{\includegraphics[width=0.45\textwidth]{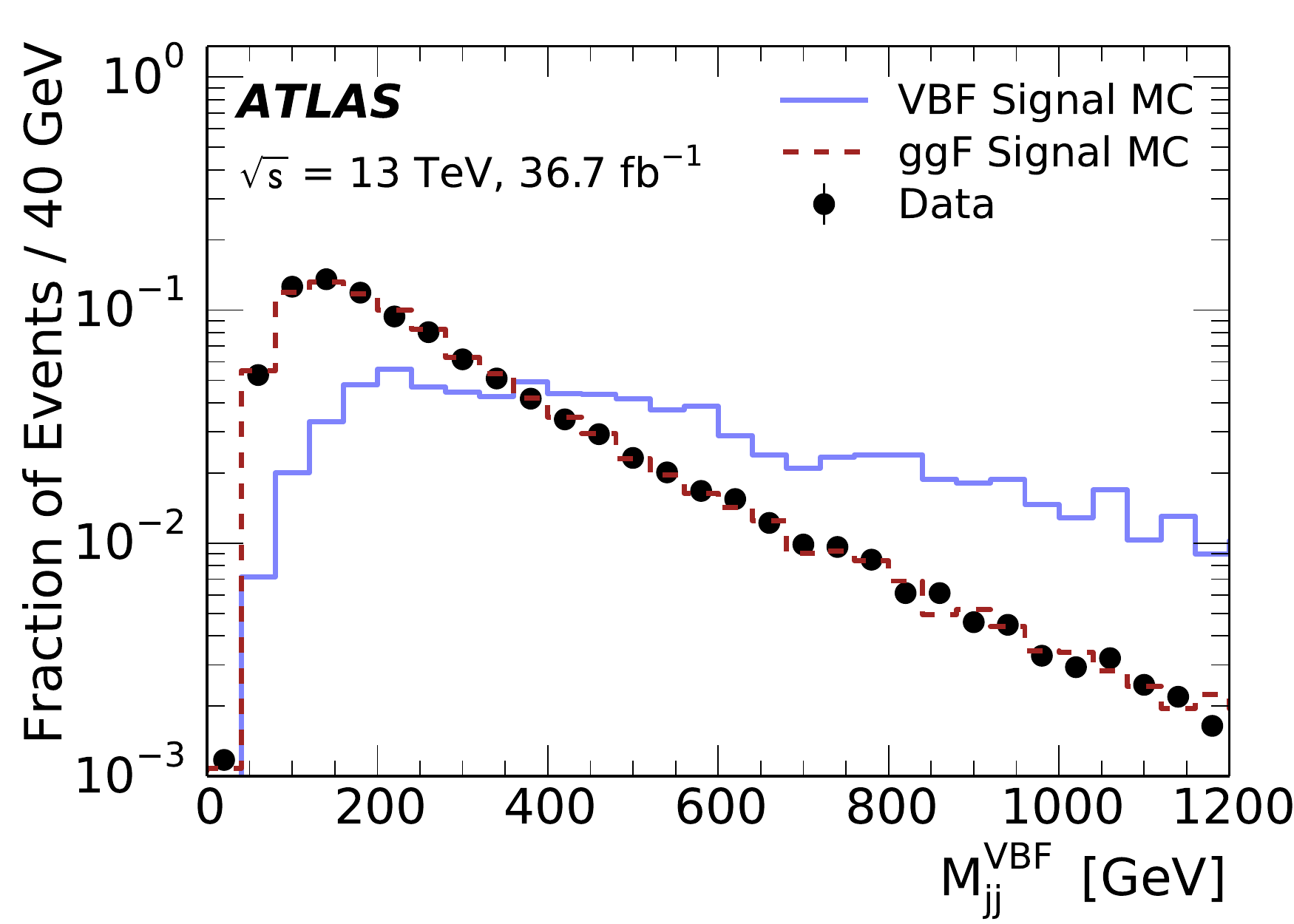}\label{fig:dista}}
  \subfloat[]{\includegraphics[width=0.45\textwidth]{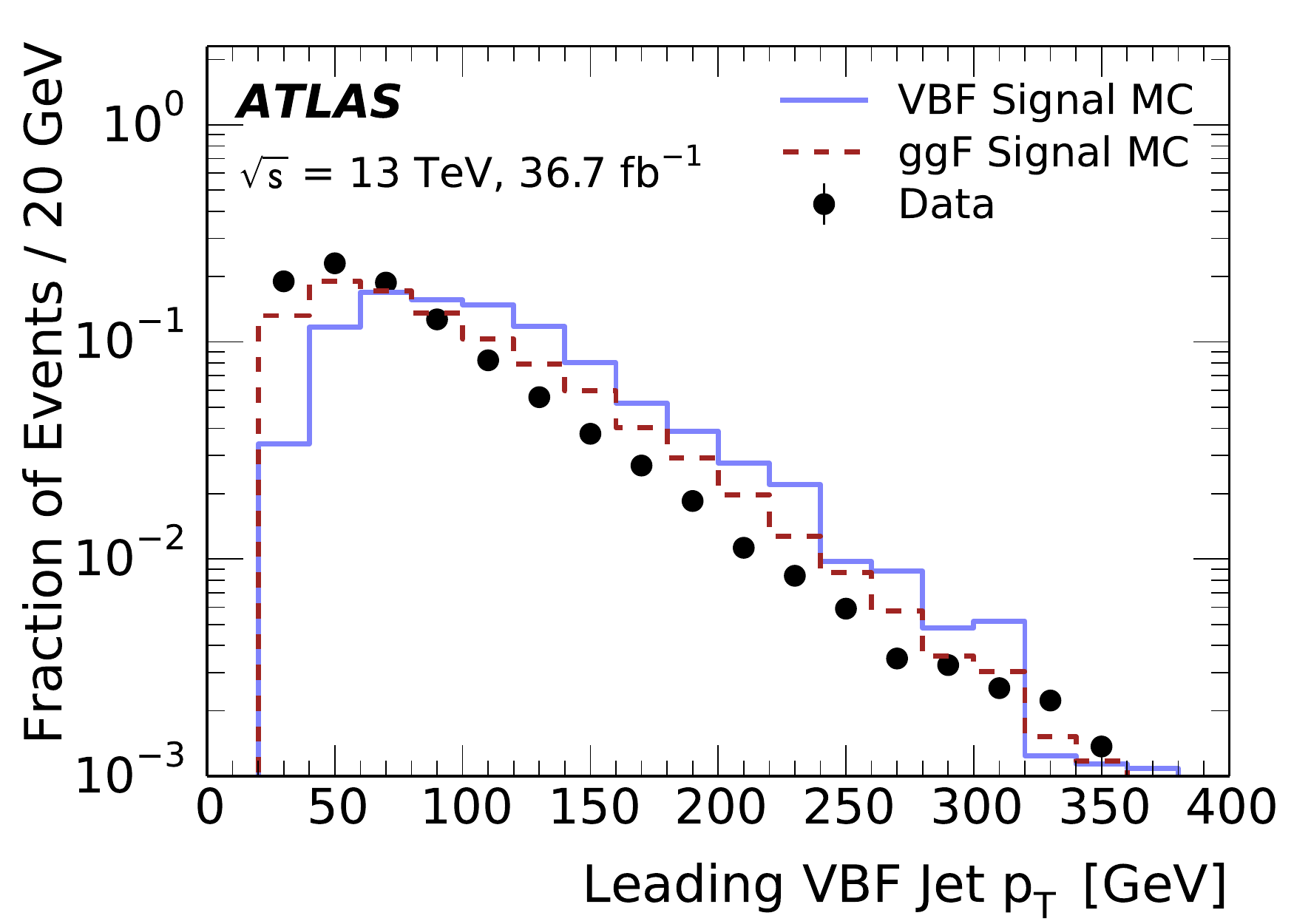}\label{fig:distb}}\\
  \subfloat[]{\includegraphics[width=0.45\textwidth]{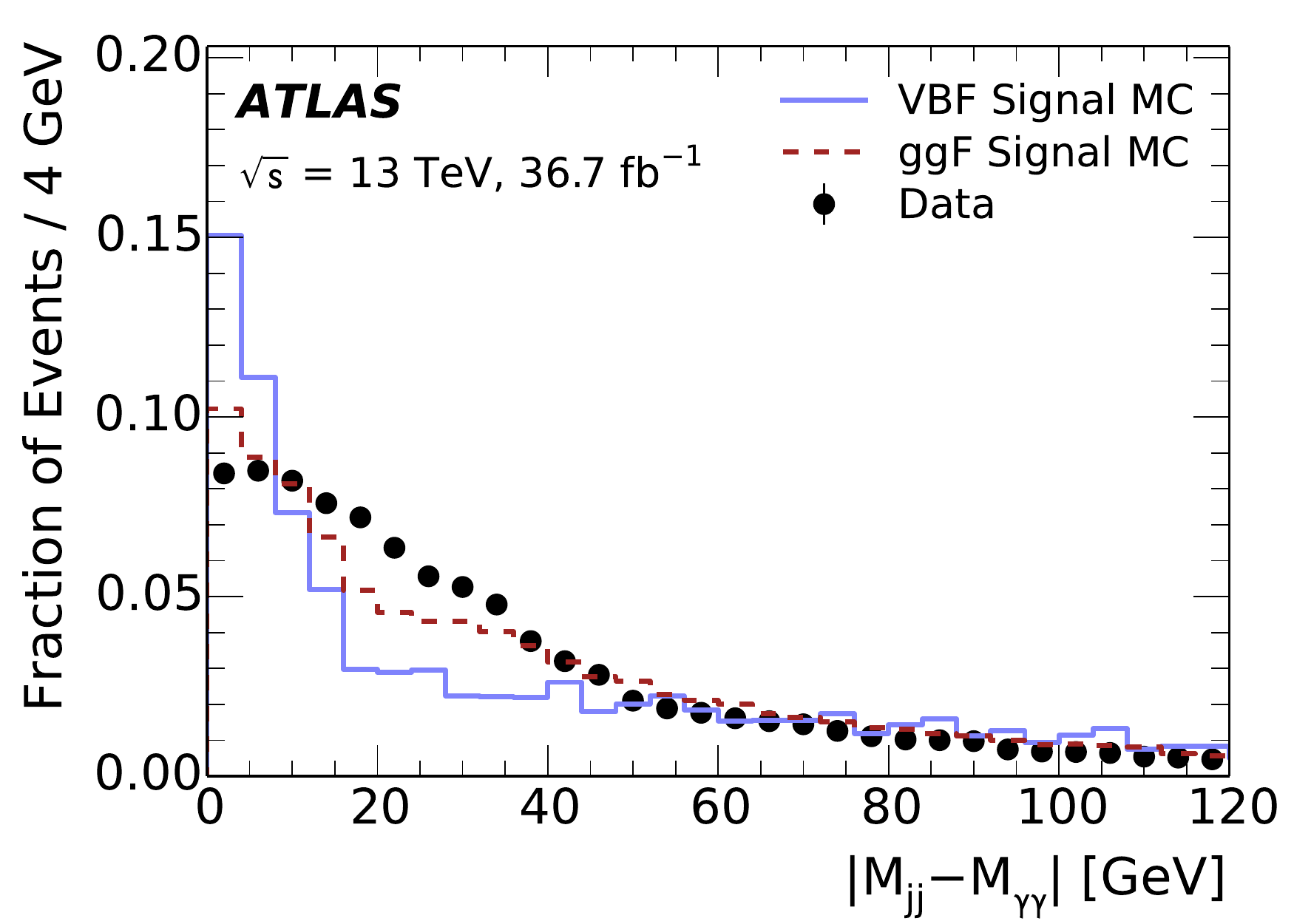}\label{fig:distc}}
  \subfloat[]{\includegraphics[width=0.45\textwidth]{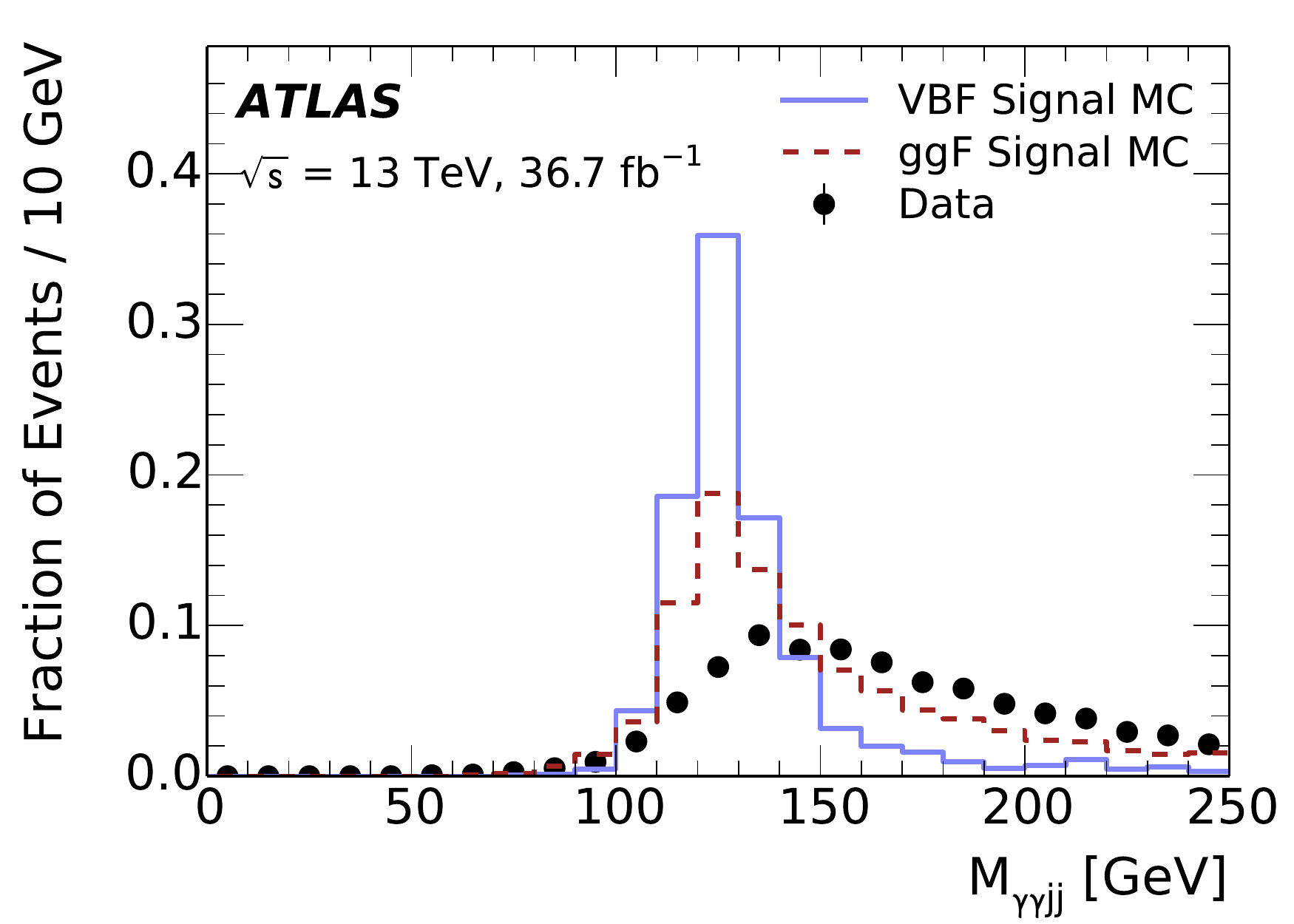}\label{fig:distd}}
  \caption{
    Distributions of kinematic observables before the requirements on \smash{$m_{jj}^\text{VBF}$}, leading VBF jet \pt{}, \smash{$m_{\gamma\gamma jj}$}
    and $|m_{jj}-m_{\gamma\gamma}|$ for:
    (a) $m_{jj}^\text{VBF}$; (b) leading VBF jet \pt{};
    (c) $|m_{jj}-m_{\gamma\gamma}|$;
    and (d) $m_{\gamma\gamma jj}$ (with the additional requirement $|m_{jj}-m_{\gamma\gamma}|< 12~\GeV{}$
    that defines the signal-enriched region).
    The quantities are shown separately for simulated signal events (with $m_a=30$ \GeV{}) produced in the VBF mode 
    and compared with those produced in the ggF mode and the observed data.
    }
  \label{fig:dist}
\end{figure}

Each event is required to have at least two photon candidates whose transverse energy requirements depend on the trigger path the event follows.
In each path the offline transverse energy requirements are designed so that the trigger selections are fully efficient.
For events passing the trigger with higher transverse energy thresholds, the leading photon is required to have $E_\text{T}>40$ \GeV{}, 
and the sub-leading photon is required to have $E_\text{T}>30$ \GeV{}. For events passing the trigger with lower thresholds, 
both the leading and sub-leading photons are required to have $E_\text{T}>27$ \GeV{}. 
For events passing both triggers, the latter selection is applied.
The invariant mass of the two leading photon candidates is denoted by $m_{\gamma\gamma}$.

In the VBF production mode, the Higgs boson is produced in association with two additional light-quark jets
with a large opening angle and a large invariant mass.
Selected events are therefore required to have at least four jets and 
the pair of jets with the highest invariant mass ($m_{jj}^\text{VBF}$) are referred to as \textit{VBF jets}.
In VBF signal events, these jets correspond to the light quarks emitting the vector bosons 55\% of the time, 
as estimated in simulation.
The VBF Higgs boson signal is further enhanced,
relative to the dominant $\gamma\gamma$+multi-jet background,
by requiring $m_{jj}^\text{VBF}$ to be greater than 500 \GeV{} and 
the \pt{} of the leading VBF jet to be greater than 60 \GeV{}.
The discrimination power of these observables can be seen in the difference in shape between the VBF signal and the 
data, shown in Figures~\ref{fig:dista} and \ref{fig:distb}.
The two remaining highest-\pt{} jets are referred to as \textit{signal jets}, with invariant mass $m_{jj}$.
The two photon candidates and the two signal jets form the Higgs boson candidate with invariant mass $m_{\gamma\gamma jj}$, 
which is required to be in the range $100 < m_{\gamma\gamma jj} < 150~\GeV{}$. 
Figure~\ref{fig:distd} shows that most of the selected signal events lie within this range, 
while the data have a broad distribution extending to higher values.

In order to take advantage of the $m_{\gamma\gamma}$ resolution of about 1.3 \GeV{} to suppress the background with $m_{\gamma\gamma}$ far from the
range of interest, five overlapping $m_{\gamma\gamma}$ regimes are defined as summarised in Table~\ref{tab:amassdiffcut}.
The boundaries of the $m_{\gamma\gamma}$ regimes are chosen so that for any value of $m_a$ 
considered in the scope of this search there is at least one regime where there is no significant signal acceptance loss
due to the $m_{\gamma\gamma}$ requirement.
For each $m_{\gamma\gamma}$ regime, the set of $m_a$ values for which this requirement causes no significant signal acceptance loss is also indicated.

\section{Background estimation}

The $\gamma\gamma$+multi-jet background consists of multi-jet events with two reconstructed photon candidates, 
originating from isolated EM radiation or from jets.
A data-driven estimation based on two-dimensional sidebands is used to predict the background yields.
The method consists of using two uncorrelated observables
to define four regions labelled A, B, C and D.

The first axis of the A/B/C/D plane separates events in regions C and D with both photons passing the \textit{Tight} requirement 
from events in regions A and B with at most one photon 
passing the \textit{Tight} requirement and at least one passing the \textit{Loose} but not the \textit{Tight} requirement. 
These regions are referred to respectively as \textit{Tight--Tight} (C and D) and \textit{Tight--Loose} (A and B). 

The second axis separates events in regions B and D, satisfying $|m_{jj}-m_{\gamma\gamma}|< x_\text{R}$, 
from events in regions A and C, satisfying $|m_{jj}-m_{\gamma\gamma}|>x_\text{R}$. 
The value $x_\text{R}$ depends on the $m_{\gamma\gamma}$ regime R to account for the degradation in resolution at higher mass.
For $H\to aa \to \gamma\gamma gg$ signal events, where the $a$ boson candidates have similar masses, the difference $|m_{jj}-m_{\gamma\gamma}|$ tends to be smaller than in the background, 
as shown in Figure~\ref{fig:distc}.
The signal events that lie outside of the range $|m_{jj}-m_{\gamma\gamma}|< x_\text{R}$ are due to poor $m_{jj}$ resolution or to incorrect assignment of the jets corresponding 
to the gluons originating from the $a$ boson decay.
Specific $x_\text{R}$ values are given in Table~\ref{tab:amassdiffcut}.
In each $m_{\gamma\gamma}$ regime, the boundary for $|m_{jj}-m_{\gamma\gamma}|$ is 0.4 times the central $m_{\gamma\gamma}$ value.
An exception is made for the lowest $m_{\gamma\gamma}$ regime, where $x_\text{R}$ is larger in order to increase the signal efficiency.

\begin{table}[t]
  \begin{center}
    \caption{
      Definition of each $m_{\gamma\gamma}$ regime, the range of $m_a$ values considered in the scope of this search with no significant signal loss acceptance due to the $m_{\gamma\gamma}$ requirement, and the corresponding boundary $x_\text{R}$ for $|m_{jj}-m_{\gamma\gamma}|$.  
    }
  \label{tab:amassdiffcut}
    {\footnotesize
  \begin{tabular}{ c c c c }
    \toprule
    $m_{\gamma\gamma}$ regime & Definition & Range of $m_a$ values & $x_\text{R}$ [\GeV{}] \\
    \midrule
    1 & $17.5$ \GeV{} $< m_{\gamma\gamma}< 27.5$ \GeV{} & $20$ \GeV{} $\le m_a \le 25$ \GeV{} & 12 \\
      2 & $22.5$ \GeV{} $< m_{\gamma\gamma}< 37.5$ \GeV{} & $25$ \GeV{} $\le m_a \le 35$ \GeV{} & 12 \\
      3 & $32.5$ \GeV{} $< m_{\gamma\gamma}< 47.5$ \GeV{} & $35$ \GeV{} $\le m_a \le 45$ \GeV{} & 16 \\
      4 & $42.5$ \GeV{} $< m_{\gamma\gamma}< 57.5$ \GeV{} & $45$ \GeV{} $\le m_a \le 55$ \GeV{} & 20 \\
      5 & $52.5$ \GeV{} $< m_{\gamma\gamma}< 65.0$ \GeV{} & $55$ \GeV{} $\le m_a \le 60$ \GeV{} & 24 \\
    \bottomrule
  \end{tabular}
    }
  \end{center}
\end{table}

Region D is expected to contain the highest contribution of signal. 
In this region, 60\% of the signal events are produced in the VBF mode and the remaining 40\% in the ggF mode.
Assuming no correlation in the background events between the two observables used to define the A/B/C/D regions,
the number of background events in the signal region D ($N^\text{bkg}_\text{D}$) is related to the number
of background events in the control regions A, B and C, denoted by $N^\text{bkg}_\text{A}$, $N^\text{bkg}_\text{B}$
and $N^\text{bkg}_\text{C}$, respectively, by the formula
\begin{align}
N^\text{bkg}_\text{D} = \frac{N^\text{bkg}_\text{B}N^\text{bkg}_\text{C}}{N^\text{bkg}_\text{A}}.
\label{eqn:closure}
\end{align}
In the following, the difference between the prediction $N^\text{bkg}_\text{D}$ and the actual background yield in region D 
is referred to as \textit{non-closure}.
The non-closure results from residual correlations between the two observables used to define the A/B/C/D regions,
and the uncertainty accounting for this effect is referred to as \textit{closure uncertainty}.
In order to quantify the non-closure, the data-driven estimation as described above is performed 
with the exception that the requirement on $m_{\gamma\gamma jj}$ is inverted.
For each $m_{\gamma\gamma}$ regime,
the closure uncertainty is defined to be 
the central value of the non-closure if it is found to be significant ($>1\sigma$) in comparison with its statistical uncertainty; 
otherwise, the statistical uncertainty of its estimate is used.

\section{Results}

The efficiency of the event selection for the inclusive $pp\to H\to aa \to \gamma\gamma gg$ signal in each of the A/B/C/D regions is shown in Table~\ref{tab:ABCD},
assuming the SM cross-section and kinematics for the ggF and VBF production modes, and the SM inclusive cross section as described in Ref.~\cite{deFlorian:2016spz};
the contribution from all other production modes is expected to be negligible.
The observed number of events in each of the A/B/C/D regions for each $m_{\gamma\gamma}$ regime is shown in Table~\ref{tab:ABCD_data}
along with the predicted background in the signal region D, taking into account the closure uncertainty. 
Due to the low event counts in each of the A/B/C/D regions,
the median expected background yield in region D estimated from pseudo-data experiments involving asymmetric Poisson uncertainties 
in the different regions slightly differs from the direct estimation from Eq.~(\ref{eqn:closure}).
No large excess is observed in region D when comparing the data yield to the background predicted from the A/B/C regions
assuming that the signal is absent in these regions.
However, given that a signal contamination is possible, a more refined procedure
taking into account signal contributions in all regions
is employed to set limits on the production rate of $H\to aa \to \gamma\gamma gg$.

\begin{table}[t]
  \begin{center}
    \caption{Efficiency of event selection on the inclusive $pp\to H\to aa \to \gamma\gamma gg$ signal, 
      assuming the SM Higgs boson production cross-section and kinematics,
      in each of the A/B/C/D regions, for different $m_a$ mass hypotheses.
      For each $m_a$ value, all $m_{\gamma\gamma}$ regimes in which there is no significant signal acceptance loss due to the $m_{\gamma\gamma}$ requirement are shown.
    }
    \label{tab:ABCD}
    \footnotesize
    \bgroup
    \def\arraystretch{1.5}
    \begin{tabular}{
        c
        c
        r@{}l
        r@{}l
        r@{}l
        r@{}l
      }
      \hline
      {$m_a$ [\GeV{}]} & {$m_{\gamma\gamma}$ regime} & \multicolumn{8}{c}{Efficiency $(\times 10^{-5})$}  \\
      & & \multicolumn{2}{c}{A} & \multicolumn{2}{c}{B} & \multicolumn{2}{c}{C} & \multicolumn{2}{c}{D}  \\
      \hline
      20 & 1 & 0.50&$^{+0.16}_{-0.14}$ & 1.2&$\pm0.4$ & 3.9&$\pm1.1$           & 6.2&$\pm1.8$           \\
      25 & 1 & 0.67&$^{+0.27}_{-0.33}$ & 2.6&$^{+0.5}_{-0.6}$ & 5.8&$\pm1.4$           & 15&$\pm4$           \\ 
      25 & 2 & 0.67&$^{+0.27}_{-0.33}$ & 2.6&$^{+0.5}_{-0.6}$ & 5.8&$\pm1.4$           & 15&$\pm4$           \\ 
      30 & 2 & 1.22&$\pm0.34$           & 3.3&$\pm0.9$          & 7.6&$^{+1.4}_{-1.6}$   & 25&$^{+5}_{-6}$   \\
      35 & 2 & 1.8&$\pm1.1$            & 2.7&$\pm1.2$           & 9.3&$\pm2.6$           & 27&$\pm6$           \\ 
      35 & 3 & 0.53&$^{+1.20}_{-0.24}$  & 4.1&$\pm1.2$           & 6.1&$^{+1.2}_{-1.6}$   & 31&$\pm7$           \\
      40 & 3 &  1.2&$\pm0.4$           & 3.3&$\pm1.0$           & 7.9&$^{+1.7}_{-2.4}$   & 26&$\pm6$           \\
      45 & 3 & 2.5&$\pm1.0$           & 4.1&$\pm1.3$           & 7.7&$^{+1.7}_{-2.0}$   & 19&$\pm5$           \\ 
      45 & 4 & 2.2&$\pm0.9$  & 4.4&$\pm1.4$           & 5.9&$^{+1.5}_{-2.2}$   & 22&$\pm5$           \\ 
      50 & 4 &  0.93&$\pm0.30$           & 4.4&$\pm1.2$           & 5.0&$^{+1.3}_{-1.0}$   & 24&$\pm5$   \\
      55 & 4 & 0.37&$\pm0.11$          & 3.3&$\pm0.9$          & 5.4&$^{+1.3}_{-1.4}$   & 21&$\pm5$           \\ 
      55 & 5 & 0.23&$\pm0.16$          & 3.6&$\pm1.0$          & 3.4&$\pm0.8$          & 24&$\pm6$           \\ 
      60 & 5 &  0.77&$^{+0.32}_{-0.30}$  & 3.9&$\pm1.0$           & 4.9&$\pm1.4$           & 23&$\pm6$           \\
      \hline
    \end{tabular}
    \egroup
  \end{center}
\end{table}

\begin{table}[t]
  \begin{center}
    \caption{Number of events observed in each of the A/B/C/D regions, 
      the relative size of the closure uncertainty considered for each $m_{\gamma\gamma}$ regime, 
      and the prediction for the number of background events in region D based on the control region yields.
      The median predicted background yield and its $\pm1\sigma$ uncertainty in region D is also shown.
      The uncertainties in the prediction account for both the Poisson fluctuations of the number of events in the control regions 
      and the closure uncertainty.
    }
    \label{tab:ABCD_data}
          {\footnotesize
            \bgroup
            \def\arraystretch{1.3}
	    \begin{tabular}{
                ccccccr@{}l
              }
	      \hline
              $m_{\gamma\gamma}$ regime &   A &   B &   C &   D  & Relative closure uncert. & \multicolumn{2}{c}{Predicted background yield}\\
	      \hline
              1 &  15 &   4 &  28 &   4 &  0.50 & \hspace{1.4cm}6&$^{+7}_{-4}$   \\
              2 &  22 &   6 &  34 &  15 &  0.32 & 8&$^{+7}_{-4}$   \\
              3 &  12 &  16 &  29 &  26 &  0.20 & 37&$^{+23}_{-14}$ \\
              4 &   8 &  12 &  19 &  38 &  0.21 & 27&$^{+22}_{-12}$ \\
              5 &   6 &  20 &  20 &  36 &  0.20 & 66&$^{+56}_{-28}$ \\
	      \hline
	    \end{tabular}
            \egroup
          }
  \end{center}
\end{table}

A likelihood function, describing both the expected background and signal, is fit to all four A/B/C/D regions simultaneously.
The free parameters of the likelihood are the numbers of signal and background events in region D, 
denoted $\mu_\text{S}$ and $\mu_\text{bkg}$ respectively, the ratio of background events expected in region B to that in region D, $\tau_\text{B}$, 
and the ratio of background events expected in region C to that in region D, $\tau_\text{C}$.
The assumption of no correlation in the total background, Eq.~(\ref{eqn:closure}), 
allows the background to be parameterised in terms of only three parameters.
The closure uncertainty, which accounts for the uncertainty due to assuming non-correlation, is included in the likelihood function by applying a Gaussian prior
to the expected number of background events in region A, $\tau_\text{B}\tau_\text{C}\mu_\text{bkg}$.
The Gaussian width is determined by the size of the closure uncertainty summarized in Table~\ref{tab:ABCD_data}.
The parameter $\mu_\text{S}$ can be expressed as the product of the total integrated luminosity, the signal cross-section 
$\sigma_H\times B(H\to aa\to \gamma\gamma gg)$, and the signal selection efficiency estimated in MC simulation
and quoted in Table~\ref{tab:ABCD}.
The signal contamination in the control regions A, B, and C is estimated from MC simulation 
and is varied coherently with $\mu_\text{S}$ in the likelihood fit.

The low number of observed events is the dominant source of uncertainty for this search.
The second largest uncertainty is due to the closure uncertainty, also statistical in nature.
Other sources of systematic uncertainty only affect the overall signal normalisation and the amount of signal contamination
in control regions A, B and C.
Dominant sources of experimental systematic uncertainty arise from the calibration and resolution of the energy of the 
jets~\cite{PERF-2016-04,PERF-2011-04}. 
Uncertainties associated with the photon energy calibration and resolution~\cite{PERF-2013-05}, as well as the photon identification and isolation
efficiencies~\cite{PERF-2013-04}, are found to be negligible. Uncertainties associated 
with the estimation of the integrated luminosity and the simulation of pile-up interactions (\textit{Lumi and Pile-up})
are found to be negligible. 
The systematic uncertainty associated with the modelling of the kinematics in signal 
events (\textit{Modelling}) is evaluated by varying the choice of scales used in the generator program and
assuming the SM Higgs boson production~\cite{Heinemeyer:2013tqa}.
It is found to be similar in size to the experimental systematic uncertainty.

Nuisance parameters corresponding to each source of uncertainty are included in the profile likelihood with Gaussian constraints.
Their effects on the estimated number of signal events $\mu_\text{S}$ are studied using Asimov~\cite{Cowan:2010js} pseudo-datasets generated
for an expected signal corresponding to the 95\% CL upper limit obtained in this search and using the values of the background 
parameters maximising the likelihood in a fit to data which assumes no signal.
Table~\ref{tab:systs} summarises the impact of each source of uncertainty varied by $\pm1\sigma$ on the maximum-likelihood estimate for $\mu_\text{S}$ in each 
of the $m_{\gamma\gamma}$ regimes for an illustrative $m_a$ hypothesis. The statistical uncertainty is the largest one for all regimes.
\begin{table}[t]
  \begin{center}
    \caption{
      Maximum fractional impact on the fitted $\mu_\text{S}$ from sources of systematic uncertainty estimated using Asimov datasets.
      The signal injected in the Asimov datasets corresponds to the observed upper limit quoted in Table~\ref{tab:limits}.
    }
    \label{tab:systs}
          {\footnotesize
	  \begin{tabular}{cccccc}
	  \hline
          &\multicolumn{5}{c}{$m_{\gamma\gamma}$ regime} \\
          Source of Uncert.   &    1  &   2  &   3  &   4  &   5  \\
            &   $m_a=20~\GeV{}$ &  $m_a=30~\GeV{}$ &  $m_a=40~\GeV{}$ &  $m_a=50~\GeV{}$ &  $m_a=60~\GeV{}$ \\
          \hline
	  Statistical          &     0.73 &     0.51 &     0.89 &     1.13 &     0.92 \\
	  Closure              &     0.44 &     0.27 &     0.39 &     0.64 &     0.89 \\
	  \hline
	  Modelling            &     0.35 &     0.34 &     0.46 &     0.42 &     0.65 \\
	  Jet                  &     0.58 &     0.38 &     0.25 &     0.90 &     0.71 \\
	  Photon               &     0.06 &     0.05 &     0.10 &     0.12 &     0.13 \\
	  Lumi and Pile-up     &     0.06 &     0.04 &     0.27 &     0.14 &     0.32 \\
	  \hline
	  \end{tabular}
          }
  \end{center}
\end{table}
The best-fit values of the parameters of the likelihood function are given in Table~\ref{tab:MLE}.
The probability that the data are compatible with the background-only hypothesis is computed for each $m_{\gamma\gamma}$ regime and no significant 
excess is observed. The smallest local $p$-value, obtained for the $m_{\gamma\gamma}$ regime 2 ($m_a\approx30~\GeV{}$), is of the order of 4\%.
\begin{table}[t]
  \begin{center}
    \caption{Maximum-likelihood fit values for each of the free parameters of the likelihood function 
      in each $m_{\gamma\gamma}$ regime for a relevant signal $m_a$ hypothesis.
      The estimated uncertainties in the fit parameters assume
      that the likelihood function is parabolic around the minimum of the fit.
    }
    \label{tab:MLE}
          {\footnotesize
	    \begin{tabular}{
                cc
                r@{}lr@{}l
                r@{}lr@{}l
              }
	      \hline
              $m_{\gamma\gamma}$ regime & $m_a$ [\GeV{}] &   \multicolumn{2}{c}{$\mu_\text{S}$} &   \multicolumn{2}{c}{$\mu_\text{bkg}$}  &   \multicolumn{2}{c}{$\tau_\text{B}$} & \multicolumn{2}{c}{$\tau_\text{C}$} \\
	      \hline
              1 & 20 &  -7&$\pm$18    & 11&$\pm$17   & 0.5 &$\pm$0.4 &  2.9&$\pm$3.1   \\
              2 & 30 &  8&$\pm$8      & 7&$\pm$6     & 0.68 &$\pm$0.32 & 4.3&$\pm$3.1   \\
              3 & 40 &  -30&$\pm$80   & 60&$\pm$70   & 0.35 &$\pm$0.19 & 0.67&$\pm$0.33 \\
              4 & 50 &  22&$\pm$28    & 16&$\pm$23   & 0.5 &$\pm$0.4 & 0.9&$\pm$1.0   \\
              5 & 60 &  -290&$\pm$260 & 340&$\pm$340 & 0.21&$\pm$0.05 & 0.24&$\pm$0.05 \\
	      \hline
	    \end{tabular}
          }
  \end{center}
\end{table}
No significant excess is observed, and an upper limit is derived at 95\% CL.            
The expected and observed exclusion limits on $\mu_\text{S}$ are given in Table~\ref{tab:limits}.
This is related to the limit on the $pp\to H\to aa \to \gamma\gamma gg$ cross-section by appropriately normalising to the measured total integrated luminosity 
and selection efficiencies relative to the inclusive signal production obtained from the ggF and VBF MC samples (Table~\ref{tab:ABCD}).
The limit is also expressed relative to the SM cross-section for the Higgs boson, shown in Figure~\ref{fig:brazil_ma}.
Within a $m_{\gamma\gamma}$ analysis regime, limits are interpolated linearly in between simulated $m_a$ values.
Finally, for each mass point, the $m_{\gamma\gamma}$ regime that yields the best expected limit is used to provide the observed exclusion limit.
The limit is calculated using a frequentist $\text{CL}_\text{s}$ calculation~\cite{Read:2002hq}. 

\begin{table}[t]
  \begin{center}
    \caption{Observed (expected) upper limits at the 95\% CL, for each of the $m_a$ values considered in the search.
      In each case, the $m_{\gamma\gamma}$ regime used to calculate the limits is also indicated.
      The limits reflect both the statistical and systematic sources of uncertainty in the fit, and the $\pm1\sigma$ widths of the expected limit distributions are also indicated.}
    \label{tab:limits}
          {\footnotesize
            \bgroup
            \def\arraystretch{1.5}
            \begin{tabular}{ccr@{}lr@{}lr@{}l}
	    \hline
            $m_{\gamma\gamma}$ regime & $m_a$ [\GeV{}] & \multicolumn{2}{c}{$\mu_\text{S}$}   & \multicolumn{2}{c}{$\sigma_H\times B(H\to aa \to \gamma\gamma gg)$ [pb]}  & \multicolumn{2}{c}{$\frac{\sigma_H}{\sigma_\text{SM}}\times B(H\to aa \to \gamma\gamma gg)$} \\
	    \hline
            1 & 20 & $10.8\Big(10.4$&$^{+4.6}_{-3.1}\Big)$   & \hspace{1.1cm}$4.8\Big(4.6$&$^{+2.1}_{-1.4}\Big)$  & \hspace{0.505cm}$0.086\Big(0.082$&$^{+0.037}_{-0.025}\Big)$ \\
            1 & 25 & $10.4\Big(10.9$&$^{+3.8}_{-2.5}\Big)$   & $1.9\Big(2.0$&$^{+0.7}_{-0.5}\Big)$  & $0.034\Big(0.036$&$^{+0.013}_{-0.008}\Big)$ \\
            2 & 25 & $28\Big(25$&$^{+8}_{-6}\Big)$   & $5.1\Big(4.7$&$^{+1.4}_{-1.1}\Big)$  & $0.092\Big(0.084$&$^{+0.026}_{-0.019}\Big)$ \\
            2 & 30 & $29\Big(24$&$^{+11}_{-6}\Big)$  & $3.1\Big(2.6$&$^{+1.1}_{-0.7}\Big)$ & $0.056\Big(0.046$&$^{+0.021}_{-0.012}\Big)$ \\
            2 & 35 & $27\Big(22$&$^{+9}_{-6}\Big)$   & $2.7\Big(2.2$&$^{+0.9}_{-0.6}\Big)$ & $0.049\Big(0.040$&$^{+0.016}_{-0.011}\Big)$  \\
            3 & 35 & $30\Big(36$&$^{+18}_{-9}\Big)$  & $2.7\Big(3.2$&$^{+1.6}_{-0.8}\Big)$ & $0.048\Big(0.057$&$^{+0.028}_{-0.014}\Big)$  \\
            3 & 40 & $31\Big(39$&$^{+19}_{-12}\Big)$ & $3.2\Big(4.0$&$^{+2.0}_{-1.2}\Big)$  & $0.058\Big(0.073$&$^{+0.035}_{-0.022}\Big)$ \\
            3 & 45 & $45\Big(53$&$^{+15}_{-20}\Big)$ & $6.3\Big(7.5$&$^{+2.1}_{-2.8}\Big)$   & $0.113\Big(0.134$&$^{+0.038}_{-0.050}\Big)$     \\
            4 & 45 & $74\Big(68$&$^{+16}_{-15}\Big)$ & $9.2\Big(8.4$&$^{+2.0}_{-1.9}\Big)$  & $0.166\Big(0.152$&$^{+0.036}_{-0.034}\Big)$ \\
            4 & 50 & $79\Big(77$&$^{+17}_{-16}\Big)$ & $9.0\Big(8.8$&$^{+2.0}_{-1.8}\Big)$  & $0.162\Big(0.159$&$^{+0.036}_{-0.032}\Big)$   \\
            4 & 55 & $73\Big(69$&$^{+11}_{-10}\Big)$  & $9.7\Big(9.1$&$^{+1.5}_{-1.2}\Big)$   & $0.173\Big(0.163$&$^{+0.026}_{-0.022}\Big)$    \\
            5 & 55 & $48\Big(59$&$^{+41}_{-19}\Big)$ & $5.5\Big(6.8$&$^{+4.7}_{-2.1}\Big)$  & $0.10\Big(0.12$&$^{+0.08}_{-0.04}\Big)$ \\
            5 & 60 & $67\Big(81$&$^{+24}_{-31}\Big)$ & $8.0\Big(9.5$&$^{+2.8}_{-3.6}\Big)$  & $0.14\Big(0.17$&$^{+0.05}_{-0.07}\Big)$   \\
	    \hline
	    \end{tabular}
            \egroup
          }
  \end{center}
\end{table}

\begin{figure}[t]
  \centering 
  {\includegraphics[width=0.9\textwidth]{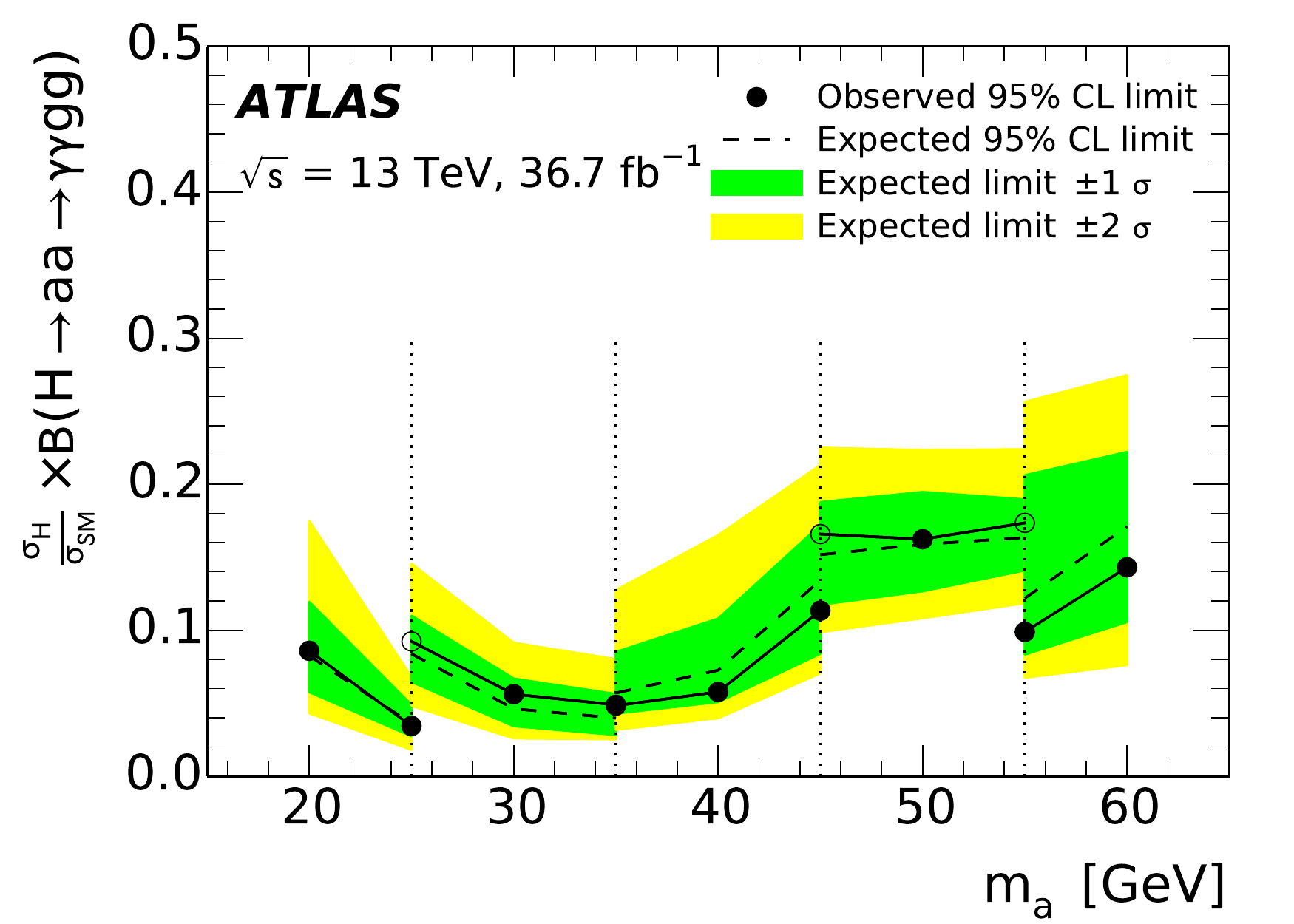}}
  \caption{The observed (solid line) and expected (dashed line) 95\% CL exclusion upper limit on 
    the \mbox{$pp\to H\to aa \to \gamma\gamma gg$} cross-section times branching ratio as a function of $m_a$,
    normalised to the SM inclusive $pp\to H$ cross-section~\cite{deFlorian:2016spz}.
  The vertical lines indicate the boundaries between the different $m_{\gamma\gamma}$ analysis regimes.
  At the boundaries, the $m_{\gamma\gamma}$ regime that yields the best expected limit is used to provide the observed exclusion limit (filled circles); the observed limit provided by the regime that yields the worse limit is also indicated (empty circles).
}
  \label{fig:brazil_ma}
\end{figure}

\FloatBarrier

\section{Conclusions}

In summary, a search for exotic decays of the Higgs boson into a pair of new (pseudo)scalar particles,
$H\to aa$, in final states with two photons 
and two jets is conducted using 36.7~\ifb{} of $pp$ collisions at $\sqrt{s}=13$ \TeV{} recorded 
with the ATLAS detector at the LHC. The search for $H\to aa \to \gamma\gamma gg$ is performed
in the mass range $20 < m_a < 60~\GeV{}$ and with additional jet requirements 
to enhance VBF-produced signal while suppressing the $\gamma\gamma$+jets background.
No significant excess of data is observed relative to the SM predictions. An upper limit
is set for the product of the production cross-section for $pp\to H$ and the branching
ratio for the decay $H\to aa\to\gamma\gamma gg$. The upper limit ranges from 3.1 pb to 9.0 pb depending
on $m_a$, and is mostly driven by the statistical uncertainties.
These results complement the previous upper limit on $H\to aa\to\gamma\gamma\gamma\gamma$ and
further constrains the BSM parameter space for exotic decays of the Higgs boson.

\section*{Acknowledgements}

% Acknowledgements for papers with collision data
% Version 14-Feb-2018

% Standard acknowledgements start here
%----------------------------------------------
We thank CERN for the very successful operation of the LHC, as well as the
support staff from our institutions without whom ATLAS could not be
operated efficiently.

We acknowledge the support of ANPCyT, Argentina; YerPhI, Armenia; ARC, Australia; BMWFW and FWF, Austria; ANAS, Azerbaijan; SSTC, Belarus; CNPq and FAPESP, Brazil; NSERC, NRC and CFI, Canada; CERN; CONICYT, Chile; CAS, MOST and NSFC, China; COLCIENCIAS, Colombia; MSMT CR, MPO CR and VSC CR, Czech Republic; DNRF and DNSRC, Denmark; IN2P3-CNRS, CEA-DRF/IRFU, France; SRNSFG, Georgia; BMBF, HGF, and MPG, Germany; GSRT, Greece; RGC, Hong Kong SAR, China; ISF, I-CORE and Benoziyo Center, Israel; INFN, Italy; MEXT and JSPS, Japan; CNRST, Morocco; NWO, Netherlands; RCN, Norway; MNiSW and NCN, Poland; FCT, Portugal; MNE/IFA, Romania; MES of Russia and NRC KI, Russian Federation; JINR; MESTD, Serbia; MSSR, Slovakia; ARRS and MIZ\v{S}, Slovenia; DST/NRF, South Africa; MINECO, Spain; SRC and Wallenberg Foundation, Sweden; SERI, SNSF and Cantons of Bern and Geneva, Switzerland; MOST, Taiwan; TAEK, Turkey; STFC, United Kingdom; DOE and NSF, United States of America. In addition, individual groups and members have received support from BCKDF, the Canada Council, CANARIE, CRC, Compute Canada, FQRNT, and the Ontario Innovation Trust, Canada; EPLANET, ERC, ERDF, FP7, Horizon 2020 and Marie Sk{\l}odowska-Curie Actions, European Union; Investissements d'Avenir Labex and Idex, ANR, R{\'e}gion Auvergne and Fondation Partager le Savoir, France; DFG and AvH Foundation, Germany; Herakleitos, Thales and Aristeia programmes co-financed by EU-ESF and the Greek NSRF; BSF, GIF and Minerva, Israel; BRF, Norway; CERCA Programme Generalitat de Catalunya, Generalitat Valenciana, Spain; the Royal Society and Leverhulme Trust, United Kingdom.

The crucial computing support from all WLCG partners is acknowledged gratefully, in particular from CERN, the ATLAS Tier-1 facilities at TRIUMF (Canada), NDGF (Denmark, Norway, Sweden), CC-IN2P3 (France), KIT/GridKA (Germany), INFN-CNAF (Italy), NL-T1 (Netherlands), PIC (Spain), ASGC (Taiwan), RAL (UK) and BNL (USA), the Tier-2 facilities worldwide and large non-WLCG resource providers. Major contributors of computing resources are listed in Ref.~\cite{ATL-GEN-PUB-2016-002}.
%----------------------------------------------

\printbibliography

\clearpage
% ATLAS Collaboration author list
% Reference date of HIGG-2017-09 is 2017-10-13
% Author list last updated on date 23-MAY-18
% Data extracted on 23-May-2018 for paper reference HIGG-2017-09
% at 8:22am
 
\begin{flushleft}
{\Large The ATLAS Collaboration}

\bigskip

M.~Aaboud$^\textrm{\scriptsize 34d}$,
G.~Aad$^\textrm{\scriptsize 99}$,
B.~Abbott$^\textrm{\scriptsize 124}$,
O.~Abdinov$^\textrm{\scriptsize 13,*}$,
B.~Abeloos$^\textrm{\scriptsize 128}$,
S.H.~Abidi$^\textrm{\scriptsize 164}$,
O.S.~AbouZeid$^\textrm{\scriptsize 143}$,
N.L.~Abraham$^\textrm{\scriptsize 153}$,
H.~Abramowicz$^\textrm{\scriptsize 158}$,
H.~Abreu$^\textrm{\scriptsize 157}$
% ...
\bigskip
\\

% ...
$^{13}$Institute of Physics, Azerbaijan Academy of Sciences, Baku; Azerbaijan.\\
% ...
$^{34}$$^{(a)}$Facult\'e des Sciences Ain Chock, R\'eseau Universitaire de Physique des Hautes Energies - Universit\'e Hassan II, Casablanca;$^{(b)}$Centre National de l'Energie des Sciences Techniques Nucleaires (CNESTEN), Rabat;$^{(c)}$Facult\'e des Sciences Semlalia, Universit\'e Cadi Ayyad, LPHEA-Marrakech;$^{(d)}$Facult\'e des Sciences, Universit\'e Mohamed Premier and LPTPM, Oujda;$^{(e)}$Facult\'e des sciences, Universit\'e Mohammed V, Rabat; Morocco.\\
% ...
$^{99}$CPPM, Aix-Marseille Universit\'e, CNRS/IN2P3, Marseille; France.\\
% ...
$^{124}$Homer L. Dodge Department of Physics and Astronomy, University of Oklahoma, Norman OK; United States of America.\\
% ...
$^{128}$LAL, Universit\'e Paris-Sud, CNRS/IN2P3, Universit\'e Paris-Saclay, Orsay; France.\\
% ...
$^{143}$Santa Cruz Institute for Particle Physics, University of California Santa Cruz, Santa Cruz CA; United States of America.\\
% ...
$^{153}$Department of Physics and Astronomy, University of Sussex, Brighton; United Kingdom.\\
% ...
$^{157}$Department of Physics, Technion, Israel Institute of Technology, Haifa; Israel.\\
$^{158}$Raymond and Beverly Sackler School of Physics and Astronomy, Tel Aviv University, Tel Aviv; Israel.\\
% ...
$^{164}$Department of Physics, University of Toronto, Toronto ON; Canada.\\
% ...
$^{a}$ Also at Borough of Manhattan Community College, City University of New York, New York City; United States of America.\\
$^{b}$ Also at Centre for High Performance Computing, CSIR Campus, Rosebank, Cape Town; South Africa.\\
$^{c}$ Also at CERN, Geneva; Switzerland.\\
$^{d}$ Also at CPPM, Aix-Marseille Universit\'e, CNRS/IN2P3, Marseille; France.\\
$^{e}$ Also at Departament de Fisica de la Universitat Autonoma de Barcelona, Barcelona; Spain.\\
% ...
$^{*}$ Deceased

\end{flushleft}

% Created with Glance <Atlas.Glance@cern.ch>

\end{document}